\documentclass[twocolumn,showpacs,superscriptaddress,amsmath,amssymb,prc]{revtex4-1}

\usepackage{graphicx}
\usepackage{dcolumn}
\usepackage{bm}
\usepackage{ifpdf}
\usepackage{epstopdf}
\usepackage{slashed}
\usepackage{amsfonts}
\usepackage{mathrsfs}
\usepackage{amssymb}
\usepackage{multirow}
\usepackage{CJK}
\usepackage{xcolor}

\begin{document}
\begin{CJK*}{GBK}{song}
\preprint{RIKEN-QHP-414}

\title{Non-relativistic expansion of Dirac equation with spherical scalar and vector potentials by similarity renormalization group}

\author{Yixin Guo}
\affiliation{Department of Modern Physics, University of Science and Technology of China, Hefei 230026, China}
\affiliation{RIKEN Nishina Center, Wako 351-0198, Japan}

\author{Haozhao Liang}
\email{haozhao.liang@riken.jp}
\affiliation{RIKEN Nishina Center, Wako 351-0198, Japan}
\affiliation{Department of Physics, Graduate School of Science, The University of Tokyo, Tokyo 113-0033, Japan}

\date{\today}

\begin{abstract}
  By following the conventional similarity renormalization group (SRG) expansion of the Dirac equation developed in [J.-Y. Guo, Phys. Rev. C \textbf{85}, 021302 (2012)], we work out the analytic expression of the ${1}/{M^4}$ order and verify the convergence of this method.
  As a step further, the reconstituted SRG method is proposed by using the re-summation technique.
  The speed of convergence of the reconstituted SRG becomes much faster than the conventional one, and the single-particle densities with the reconstituted SRG are also almost identical to the exact values.
\end{abstract}

\maketitle

\section{Introduction}

In atomic physics, it is well known that the fine structure describes the splitting of the spectral lines of atoms due to the electron spin and the relativistic corrections to the non-relativistic Schr\"odinger equation.
The leading-order corrections, including the relativistic correction to the kinetic energy, the correction due to the spin-orbit coupling, and the Darwin term, can be derived from the standard non-relativistic expansion of the Dirac equation by using the Taylor series \cite{Sommerfeld1940Naturwissenschaften28.417--423, zelevinsky2011quantum}.
In nuclear physics, since the 1970s, the density function theory (DFT) in both the non-relativistic and relativistic frameworks has achieved great successes in describing and understanding the ground-state and excited-state properties of thousands of nuclei in a microscopic and self-consistent way.
The non-relativistic expansion of the Dirac equation is considered to be a potential bridge for the connection between these two frameworks \cite{Reinhard1989Rep.Prog.Phys.52.439--514, Bender2003Rev.Mod.Phys.75.121--180}.
Recently, the non-relativistic expansion of the Dirac equation also shows promising applications for investigating the origin of the pseudospin symmetry (PSS) and its breaking mechanism \cite{Guo2012Phys.Rev.C85.021302}.

The PSS is the quasi-degeneracy phenomenon between two single-nucleon states with the quantum numbers $(n-1,l + 2,j = l + 3/2)$ and $(n,l,j = l + 1/2)$
\cite{Arima1969Phys.Lett.B30.517--522, Hecht1969Nucl.Phys.A137.129--143}.
Various phenomena in nuclear structure have been discussed based on this concept, such as superdeformation \cite{Dudek1987Phys.Rev.Lett.59.1405--1408},
identical bands \cite{Nazarewicz1990Phys.Rev.Lett.64.1654--1657,Zeng1991Phys.Rev.C44.R1745--R1748}, pseudospin partner bands \cite{Xu2008Phys.Rev.C78.064301,Hua2009Phys.Rev.C80.034303}, and so on.
See also reviews \cite{Ginocchio2005Phys.Rep.414.165--261, Liang2015Phys.Rep.570.1--84} and the references therein.
In 1997, the PSS was shown to be a symmetry of the Dirac Hamiltonian, where the pseudo-orbital angular momentum $\tilde l$ is nothing but the orbital angular momentum of the lower component of the Dirac spinor \cite{Ginocchio1997Phys.Rev.Lett.78.436--439}.
In addition, the equality in magnitude but difference in sign of the scalar potential $S(r)$ and vector potential $V(r)$ was suggested as the exact PSS limit by reducing the Dirac equation to a Schr\"odinger-like equation \cite{Ginocchio1997Phys.Rev.Lett.78.436--439}.
A more general condition $d(S + V)/dr = 0$ was proposed \cite{Meng1998Phys.Rev.C58.R628--R631}, and it can be approximately satisfied in exotic nuclei with highly diffuse potentials \cite{Meng1999Phys.Rev.C59.154--163}.
Along with this direction, a conventional way is to reduce the Dirac equation to a non-relativistic Schr\"odinger-like equation for the upper or lower component of the Dirac spinor.
However, the effective Hamiltonian thus obtained is not Hermitian, since the upper- or lower-component wave functions alone, as the solutions of the Schr\"odinger-like equation, are not orthogonal to each other.

In 2012, Guo \cite{Guo2012Phys.Rev.C85.021302} proposed a novel method to perform the non-relativistic expansion of the Dirac equation by using the similarity renormalization group (SRG).
With this method, the Dirac Hamiltonian can be transformed into a diagonal form, i.e., the eigenequations for the upper and lower components of the Dirac spinors are decoupled.
The non-relativistic reduced Hamiltonian thus obtained is Hermitian, and it can be expanded into the series in terms of ${1}/{M^i}$ (with $M$ the bare mass of nucleon).
It was also proven that, by using the SRG method, the relativistic correction to the kinetic energy appears in the order of $1/M^3$, and the correction due to the spin-orbit coupling and the Darwin term appear in the order of $1/M^2$, respectively.
In short, SRG provides a systematic way to derive all terms up to a given order.

Based on the non-relativistic expansion with SRG, the study of PSS was carried out not only in spherical \cite{Li2013Phys.Rev.C87.044311,
Shi2014Phys.Rev.C90.034318, Xu2015Eur.Phys.J.A51.81, Chen2016Sci.ChinaPhys.Mech.Astro.59.682011, Sun2017Int.J.Mod.Phys.E26.1750025} but also in deformed \cite{Guo2014Phys.Rev.Lett.112.062502, Li2015Phys.Rev.C91.024311} nuclei.
As a step further, by using the non-relativistic reduced Hamiltonian by SRG, the origin of PSS and its breaking mechanism were investigated quantitatively in the scheme of supersymmetry quantum mechanics \cite{Liang2013Phys.Rev.C87.014334, Shen2013Phys.Rev.C88.024311, Liang2016Phys.Scr.91.083005}.

However, different from the atomic systems, in the nuclear systems the most slowly convergent series is that in the powers of $S/M$, whose value is approximately $1/3$, as typically $ S \sim - 350$~MeV and $M = 939$~MeV.
As a result, even for the $1/M^3$ order, the absolute value of its contribution to the single-particle energy is up to $3$~MeV.
The remaining discrepancy between the corresponding single-particle energies and the exact ones is still around or more than $1$~MeV for most single-particle states \cite{Guo2012Phys.Rev.C85.021302}.
In order to verify the convergence of the SRG method, it is necessary to work out the non-relativistic expansion up to higher orders and evaluate their contributions to the single-particle energy.

In this work, we will perform the non-relativistic expansion of the Dirac equation by using the SRG method up to the ${1}/{M^4}$ order for the first time.
Beyond that, we will propose a re-summation method, where the expansion is written in terms of ${1}/{{M^\ast}^i}$ (with $M^\ast(r) = M +S(r)$ the Dirac effective mass of nucleon).
It will be proven that the new method provides much faster convergence than the conventional expansion for both single-particle energies and densities.

This paper is organized as follows.
In Sec.~\ref{sec:1}, the theoretical framework of the conventional SRG method will be recalled, and the expansion up to the ${1}/{M^4}$ order will be shown.
The theoretical framework for the novel reconstituted SRG method with re-summation will be introduced in Sec.~\ref{sec:new}.
The numerical details for solving the radial Schr\"odinger equation in coordinate space and the results for the single-particle energies and densities will be presented in Sec.~\ref{sec:III}.
Finally, a summary and perspectives will be given in Sec.~\ref{sec:IV}.

\section{Theoretical Framework}

\subsection{Conventional SRG method}\label{sec:1}

In the relativistic scheme, the Dirac Hamiltonian with the scalar $S$ and vector $V$ potentials for nucleons reads
\begin{equation}
H=\boldsymbol{\alpha} \cdot \mathbf{p}+\beta(M+S)+V,
\end{equation}
where $\alpha$ and $\beta$ are the Dirac matrices, $M$ is the mass of nucleon.

Following Wegner's formulation and the technique of SRG \cite{Wegner1994Ann.Phys.Berlin506.77--91,Bylev1998Phys.Lett.B428.329--333},
the Hamiltonian is transformed by a unitary operator $U(l)$ as
\begin{equation}
H(l)=U(l)HU^\dagger(l),\quad H(0)=H,
\end{equation}
with a flow parameter $l$.
Further with the anti-Hermitian generator $\eta(l)=\frac{\textrm{d}U(l)}{\textrm{d}l}U^\dagger(l)$, the Hamiltonian flow equation is obtained from the differential of the initial Hamiltonian $H$ as
\begin{equation}\label{eq:1}
    \frac{\textrm{d}H(l)}{\textrm{d}l}=[\eta(l),H(l)].
\end{equation}
It is convenient and appropriate to choose $\eta(l)$ in the form of
\begin{equation}\label{eq:2}
    \eta(l)=[\beta M,H(l)]
\end{equation}
to transform Dirac Hamiltonian into a diagonal form.

In order to solve Eq.~(\ref{eq:1}), the technique in Ref.~\cite{Bylev1998Phys.Lett.B428.329--333} was adopted in Ref.~\cite{Guo2012Phys.Rev.C85.021302}.
According to the commutation and anti-commutation relations with respect to $\beta$, the Hamiltonian is divided into two parts:
\begin{equation}\label{eq:3}
    H(l)=\varepsilon(l)+o(l),
\end{equation}
where $\varepsilon(l)$ is an even operator or a diagonal part, and $o(l)$ is an odd operator or an off-diagonal part.
They satisfy $[\varepsilon, \beta] = 0$ and $\{o, \beta\}$ = 0, respectively.

Combined with Eqs.~(\ref{eq:2}) and (\ref{eq:3}), Eq.~(\ref{eq:1}) is transformed into
\begin{subequations}
\begin{align}
\frac{\textrm{d}\varepsilon(l)}{\textrm{d}l}&=4M\beta o^2(l),\label{eq:s1} \\
\frac{\textrm{d}o(l)}{\textrm{d}l}&=2M\beta[o(l),\varepsilon(l)],\label{eq:s2}
\end{align}
\end{subequations}
and the initial conditions are
\begin{equation}
\varepsilon(0)=\beta(M+S)+V,\qquad o(0)=\boldsymbol{\alpha} \cdot \mathbf{p}.
\end{equation}

Equations~(\ref{eq:s1}) and (\ref{eq:s2}) can be solved perturbatively in ${1}/{M}$ \cite{Bylev1998Phys.Lett.B428.329--333}.
It is convenient to introduce a dimensionless flow parameter $\lambda=lM^2$. The expansions of ${\varepsilon(\lambda)}/{M}$ and ${o(\lambda)}/{M}$ are then done as follows:
\begin{subequations}
\begin{align}
\frac{\varepsilon(\lambda)}{M}&=\sum^\infty_{i=0}\frac{\varepsilon_i(\lambda)}{M^i},\label{eq:s3}\\
\frac{o(\lambda)}{M}&=\sum^\infty_{j=1}\frac{o_i(\lambda)}{M^i}.\label{eq:s4}
\end{align}
\end{subequations}

To differentiate Eqs.~(\ref{eq:s3}) and (\ref{eq:s4}) gives \cite{Guo2012Phys.Rev.C85.021302}
\begin{subequations}
\begin{align}
\frac{\textrm{d}\varepsilon_n(\lambda)}{\textrm{d}\lambda} &= 4\beta\sum^{n-1}_{k=1}o_k(\lambda)o_{n-k}(\lambda),\label{eq:s5}\\
\frac{\textrm{d}o_n(\lambda)}{\textrm{d}\lambda} &= -4o_n(\lambda)+2\beta\sum^{n-1}_{k=1}[o_k(\lambda),\varepsilon_{n-k}(\lambda)].\label{eq:s6}
\end{align}
\end{subequations}
The solutions of Eqs.~(\ref{eq:s5}) and (\ref{eq:s6}) are obtained as \cite{Guo2012Phys.Rev.C85.021302}
\begin{subequations}
\begin{align}
\varepsilon_n(\lambda)&=\varepsilon_n(0)+4\beta\int_0^\lambda\sum^{n-1}_{k=1}o_k(\lambda')o_{n-k}(\lambda')\,\textrm{d}\lambda',\label{eq:s7}\\
o_n(\lambda)&=o_n(0)e^{-4\lambda}\nonumber\\
&\quad+2\beta e^{-4\lambda}\int_0^\lambda\sum^{n-1}_{k=1}[e^{4\lambda'}o_k(\lambda'),\varepsilon_{n-k}(\lambda')]\,\textrm{d}\lambda',\label{eq:s8}
\end{align}
\end{subequations}
with the initial conditions,
\begin{align}\label{eq:s9}
&\varepsilon_0(0)=\beta,\quad  \varepsilon_1(0)=\beta S+V,\quad  \varepsilon_n(0)=0\quad  \mbox{if}\quad  n\geq2,\nonumber\\
&o_1(0)=\boldsymbol{\alpha} \cdot \mathbf{p},\quad o_n(0)=0\quad \mbox{if}\quad  n\geq2.
\end{align}

Therefore, when $\lambda \rightarrow \infty$ the diagonalized Dirac operator up to the $1/M^3$ order is easy to be obtained as \cite{Guo2012Phys.Rev.C85.021302}
\begin{align}\label{eq:A1}
\varepsilon(\infty)
=&\,M\varepsilon_0(\infty)+\varepsilon_1(\infty)+\frac{\varepsilon_2(\infty)}{M}+\frac{\varepsilon_3(\infty)}{M^2}\nonumber\\
&+\frac{\varepsilon_4(\infty)}{M^3}+\cdots\nonumber\\
=&\,M\varepsilon_0(0)+\varepsilon_1(0)+\frac{1}{2M}\beta o_1^2(0)\nonumber\\
&+\frac{1}{8M^2}[[o_1(0),\varepsilon_1(0)],o_1(0)]\nonumber\\
&+\frac{1}{32M^3}\beta(-4o_1^4(0)+[[o_1(0),\varepsilon_1(0)],\varepsilon_1(0)]o_1(0)\nonumber\\
&+o_1(0)[[o_1(0),\varepsilon_1(0)],\varepsilon_1(0)]\nonumber\\
&-2[o_1(0),\varepsilon_1(0)][o_1(0),\varepsilon_1(0)])+\cdots
\end{align}
Furthermore, the results of the next order, i.e., the ${1}/{M^4}$ order, can also be worked out carefully as
\begin{align}
\frac{\varepsilon_5(\infty)}{M^4}=&\frac{1}{128M^4}(-9[[o_1(0),\varepsilon_1(0)],o_1^3(0)]\nonumber\\
&+3[o_1(0),\varepsilon_1(0)]^2\varepsilon_1(0)+3\varepsilon_1(0)[o_1(0),\varepsilon_1(0)]^2\nonumber\\
&-6[o_1(0),\varepsilon_1(0)]\varepsilon_1(0)[o_1(0),\varepsilon_1(0)]\nonumber\\
&+3[o_1(0)[o_1(0),\varepsilon_1(0)]o_1(0),o_1(0)]\nonumber\\
&+[[[[o_1(0),\varepsilon_1(0)],\varepsilon_1(0)],\varepsilon_1(0)],o_1(0)]).
\end{align}

For the systems with spherical symmetry, the corresponding Dirac equation reads
\begin{equation}\label{eq:Dirac}
\left( \begin{array}{cc}
\Sigma(r)+M & -\frac{\textrm{d}}{\textrm{d}r}+\frac{\kappa}{r}\\
\frac{\textrm{d}}{\textrm{d}r}+\frac{\kappa}{r} & \Delta(r)-M
\end{array}
\right )
\left( \begin{array}{c}
G(r) \\ F(r)
\end{array}
\right )
=
E
\left( \begin{array}{c}
G(r) \\ F(r)
\end{array}
\right ),
\end{equation}
where $\kappa$ is a good quantum number defined as $\kappa=\mp~(j+{1}/{2})$ for $j=l\pm{1}/{2}$, and $\Sigma(r) = V(r) + S(r)$ and $\Delta(r) = V(r) - S(r)$ are the sum of and the difference between the vector and scalar potentials, respectively.
The single-particle energy $E = \varepsilon +M$ including the mass of nucleon.

The initial conditions in Eq.~(\ref{eq:s9}) are then written as
\begin{equation}\label{eq:pi}
\varepsilon_1(0)={\left( \begin{array}{cc}
\Sigma(r) & 0\\
0 & \Delta(r)
\end{array}
\right )},\quad
o_1(0)={\left( \begin{array}{cc}
0 &-\frac{\textrm{d}}{\textrm{d}r}+\frac{\kappa}{r}\\
\frac{\textrm{d}}{\textrm{d}r}+\frac{\kappa}{r}& 0
\end{array}
\right )}.
\end{equation}
At the end of the flow, the Dirac Hamiltonian is transformed into a diagonal form as $o(\infty) = 0$ and
\begin{equation}
\varepsilon(\infty)=
{\left( \begin{array}{cc}
\mathcal{H}^{\rm (F)}+M & 0\\
0 & \mathcal{H}^{\rm (D)} - M
\end{array}
\right )}.
\end{equation}
As a result, the eigenequations for the upper and lower components of the Dirac spinors are decoupled.

Hereafter, we will focus on the single-particle states in the Fermi sea, which correspond to their counterparts in the non-relativistic framework.
Therefore, $\mathcal{H}^{\rm (F)}$ will be investigated in detail, and its superscript will be omitted when there is no confusion.

In Ref.~\cite{Guo2012Phys.Rev.C85.021302}, the expansion of $\mathcal{H}$ was given up to the $1/M^3$ order as
\begin{subequations}
\begin{align}
\mathcal{H}_0=&\Sigma(r),\label{eq:p0}\\
\mathcal{H}_1=&\frac{1}{2M}p^2,\label{eq:p1}\\
\mathcal{H}_2=&\frac{1}{8M^2}(-4Sp^2+4S'\frac{\textrm{d}}{\textrm{d}r}-2\frac{\kappa}{r}\Delta'+\Sigma''),\label{eq:p2}\\
\mathcal{H}_3=&\frac{1}{32M^3}(-4p^4+16S^2p^2-32SS'\frac{\textrm{d}}{\textrm{d}r}\nonumber\\
&-8S\Sigma''+16S\Delta'\frac{\kappa}{r}-2\Sigma'^2+4\Sigma'\Delta'),\label{eq:p3}
\end{align}
\end{subequations}
where
\begin{equation}
p^2=-\frac{\textrm{d}^2}{\textrm{d}r^2}+\frac{\kappa(\kappa+1)}{r^2}
\end{equation}
and
\begin{align}
p^4=&\frac{\textrm{d}^4}{\textrm{d}r^4}-2\frac{\kappa(\kappa+1)}{r^2}\frac{\textrm{d}^2}{\textrm{d}r^2}
+4\frac{\kappa(\kappa+1)}{r^3}\frac{\textrm{d}}{\textrm{d}r}\nonumber\\
&+\frac{\kappa(\kappa+1)(\kappa+3)(\kappa-2)}{r^4}.
\end{align}

In order to verify the convergence of the SRG approach, here the results of the ${1}/{M^4}$ order are worked out as
\begin{align}\label{eq:order4p}
\mathcal{H}_{4}=&\frac{1}{128M^4}
\Big\{48Sp^4-96S'p^2\frac{\textrm{d}}{\textrm{d}r}\nonumber\\
&+[24\Delta'\frac{\kappa}{r}-24(\Sigma''+3S'')-64S^3]p^2\nonumber\\
&+[24\Delta'\frac{\kappa}{r^2}-24\Delta''\frac{\kappa}{r}+24(S'''+\Sigma''')+192S^2S']\frac{\textrm{d}}{\textrm{d}r}\nonumber\\
&+[-12(\Sigma'-12S')\frac{\kappa(\kappa+1)}{r^3}-24\Delta'\frac{\kappa}{r^3}\nonumber\\
&+12(\Sigma''+4S'')\frac{\kappa(\kappa+1)}{r^2}+24\Delta''\frac{\kappa}{r^2}\nonumber\\
&-12\Delta'''\frac{\kappa}{r}-96S^2\Delta'\frac{\kappa}{r}+9\Sigma''''+48S^2\Sigma''\nonumber\\
&+24S\Sigma'(\Sigma'-2\Delta')]\Big\}.
\end{align}

It is noted that operators with higher-order derivatives appear from $\mathcal{H}_3$, and thus the eigenequation containing up to the second derivatives reads
\begin{equation}\label{eq:Sch1}
  \left[\mathcal{H}_0 + \mathcal{H}_1 + \mathcal{H}_2\right] \varphi_k(r) = \varepsilon_k \varphi_k(r).
\end{equation}
In the following discussions, the eigenequation~(\ref{eq:Sch1}) will be solved, and the higher-order terms $\mathcal{H}_3$ and $\mathcal{H}_4$ will be calculated by the perturbation theory.

\subsection{Reconstituted SRG method}\label{sec:new}

In Sec.~\ref{sec:1}, the conventional SRG approach introduced in Ref.~\cite{Guo2012Phys.Rev.C85.021302} has been followed.
However, its speed of convergence is rather slow, because the most slowly convergent series is that in terms of the power of ${S}/{M}$, whose value is approximately ${1}/{3}$.
As a result, the biggest contribution to the single-particle energy at each $1/M^i$ order comes from the term $\frac{(-S)^{i-1}} { 2M^i} p^2$, which will be shown numerically in Sec.~\ref{sec:IIIA}.
This observation reminds us the idea of re-summation, which is widely used in the studies such as the Brueckner theory \cite{Brueckner1954Phys.Rev.95.217--228, Day1967Rev.Mod.Phys.39.719--744}, and so on.

In the present case, there is only one term in $\mathcal{H}_1$,
\begin{align}
\frac{1}{2M}p^2,
\end{align}
and it can be seen that this term is accompanied by its family in the higher orders, i.e.,
\begin{align}
\frac{1}{2M}p^2-\frac{S}{2M^2}p^2+\frac{S^2}{2M^3}p^2-\frac{S^3}{2M^4}p^2+\cdots
\end{align}
Indeed, we can easily sum this series up to the infinite order as
\begin{align}
&\frac{1}{2M}p^2-\frac{S}{2M^2}p^2+\frac{S^2}{2M^3}p^2-\frac{S^3}{2M^4}p^2+\cdots\\\nonumber
=&\frac{1}{2(M+S)}p^2\equiv\frac{1}{2M^\ast}p^2.
\end{align}
Here the effective mass $M^\ast$ is nothing but the well-known Dirac mass in the relativistic scheme.

In the next order, ${1}/{M^2}$, we have additional terms that do not belong to the above family,
\begin{align}
\frac{S'}{2M^2}\frac{\textrm{d}}{\textrm{d}r}-\frac{\kappa}{r}\frac{\Delta'}{4M^2}+\frac{\Sigma''}{8M^2}.
\end{align}
These terms are also accompanied by their family in the higher orders.
For example,
\begin{equation}
\frac{S'}{2M^2}\frac{\textrm{d}}{\textrm{d}r}
-\frac{2SS'}{2M^3}\frac{\textrm{d}}{\textrm{d}r}
+\frac{3S^2S'}{2M^4}\frac{\textrm{d}}{\textrm{d}r}
-\cdots
=\frac{S'}{2{M^\ast}^2}\frac{\textrm{d}}{\textrm{d}r}.
\end{equation}
The same is also true for the new terms in the ${1}/{M^3}$ order,
\begin{align}
\frac{\Sigma'^2}{16M^3}-\frac{S'\Sigma'}{4M^3}-\frac{1}{8M^3}p^4.
\end{align}
For example,
\begin{equation}
-\frac{1}{8M^3}p^4+\frac{3S}{8M^4}p^4-\cdots=-\frac{1}{8{M^\ast}^3}p^4.
\end{equation}

In short, by replacing the bare mass $M$ by the Dirac mass $M^\ast$, we not only obtain the non-relativistic expansion up to a certain order but also sum up the terms that belong to their families up to the infinite order.

Up to the ${1}/{M^3}$ order, we have the new expression
\begin{align}\label{eqh1}
\mathcal{H}= &\,\Sigma +\frac{1}{2M^\ast}p^2 +\frac{S'}{2{M^\ast}^2}\frac{\textrm{d}}{\textrm{d}r} -\frac{\Delta'}{4{M^\ast}^2}\frac{\kappa}{r} +\frac{\Sigma''}{8{M^\ast}^2}\nonumber\\
&+\frac{\Sigma'^2}{16{M^\ast}^3} -\frac{S'\Sigma'}{4{M^\ast}^3} -\frac{1}{8{M^\ast}^3}p^4+\cdots
\end{align}

Let us then discuss about the Hermitian properties of the Hamiltonian. For example, on the one hand, now $M^\ast$ is not a constant, and the differential operator $p^2$ is acting only to the right-hand side. Thus, the term $\frac{1}{2M^\ast}p^2$ alone in the Hamiltonian is not Hermitian.
On the other hand, the following operator is Hermitian, which is nothing but
\begin{align}
 -\frac{\textrm{d}}{\textrm{d}r}\frac{1}{2M^\ast}\frac{\textrm{d}}{\textrm{d}r}+\frac{\kappa(\kappa+1)}{2M^\ast r^2}=\frac{1}{2M^\ast}p^2+\frac{S'}{2{M^\ast}^2}\frac{\textrm{d}}{\textrm{d}r}.
\end{align}
In other words, in the expansion, we cannot just arbitrarily truncate up to a certain order, such as ${1}/{M^\ast}$, instead we should keep the corresponding terms in higher orders in order to make the truncated Hamiltonian Hermitian, such as those two terms above, although the second term belongs to the next order. Now let us see which terms we should keep together with $-\frac{1}{8{M^\ast}^3}p^4$. The corresponding Hermitian operator reads
\begin{align}
-p^2\frac{1}{8{M^\ast}^3}p^2
=&-\frac{1}{8{M^\ast}^3}p^4-\frac{3S'}{4{M^\ast}^4}p^2\frac{\textrm{d}}{\textrm{d}r}\nonumber \\
&+(-\frac{3S''}{8{M^\ast}^4}+\frac{3S'^2}{2{M^\ast}^5})p^2\nonumber \\
&+\frac{3S'}{2{M^\ast}^4}\frac{\kappa(\kappa+1)}{r^3}.
\end{align}

Finally, the Hamiltonian in the new method, which we call the reconstituted SRG expansion, is obtained as
\begin{subequations}
\begin{align}
\tilde{\mathcal{H}}_0&=\Sigma(r),\\
\tilde{\mathcal{H}}_1&=-\frac{\textrm{d}}{\textrm{d}r}\frac{1}{2M^\ast}\frac{\textrm{d}}{\textrm{d}r}+\frac{\kappa(\kappa+1)}{2M^\ast r^2},\\
\tilde{\mathcal{H}}_2&=-\frac{\Delta'}{4{M^\ast}^2}\frac{\kappa}{r}+\frac{\Sigma''}{8{M^\ast}^2},\\
\tilde{\mathcal{H}}_3&=-p^2\frac{1}{8{M^\ast}^3}p^2 +\frac{\Sigma'^2}{16{M^\ast}^3}-\frac{S'\Sigma'}{4{M^\ast}^3}.\label{eqn3}
\end{align}
\end{subequations}

Similar to the conventional expansion, operators with higher-order derivatives appear from $\tilde{\mathcal{H}}_3$, and thus the eigenequation containing up to the second derivatives reads
\begin{equation}\label{eq:Sch2}
  \left[\tilde{\mathcal{H}}_0 + \tilde{\mathcal{H}}_1 + \tilde{\mathcal{H}}_2\right] \tilde{\varphi}_k(r) = \tilde{\varepsilon}_k \tilde{\varphi}_k(r).
\end{equation}
In the following discussions, the eigenequation~(\ref{eq:Sch2}) will be solved, and the higher-order term $\tilde{\mathcal{H}}_3$ will be calculated by the perturbation theory.

\section{Results and Discussion}\label{sec:III}

In order to figure out the differences and improvements to the results and discussion in Ref.~\cite{Guo2012Phys.Rev.C85.021302}, we use the Woods-Saxon potentials for $\Sigma(r)$ and $\Delta(r)$, i.e., $\Sigma(r)=\Sigma_0f(a_0,r_0,r)$ and $\Delta(r)=\Delta_0f(a_0,r_0,r)$ with
\begin{equation}
f(a_0,r_0,r)=\frac{1}{1+e^{\frac{r-r_0}{a_0}}},
\end{equation}
where the parameters are the same as those in Ref.~\cite{Guo2012Phys.Rev.C85.021302}, i.e., $\Sigma_0=-66.0$~MeV, $\Delta_0 = 650.0$~MeV, $r_0= 7.0$~fm, and $a_0= 0.6$~fm, which are the typical values for the nucleus $^{208}\textrm{Pb}$.
The mass of nucleon is taken as $M=939.0$~MeV.

The Dirac equation~(\ref{eq:Dirac}) is solved in coordinate space by the shooting method \cite{Meng1998Nucl.Phys.A635.3--42} within a spherical box with
radius $R_{\rm box}= 20$~fm and mesh size $\textrm{d}r = 0.05$~fm.
The single-particle energies and densities thus obtained will serve as benchmarks, labelled as Exact in the tables and figures.
The non-relativistic equations~(\ref{eq:Sch1}) and (\ref{eq:Sch2}) are also solved in coordinate space by the shooting method with the same box and mesh sizes.

\subsection{Single-particle energy with conventional SRG}\label{sec:IIIA}

\begin{table}
  \caption{Eigenenergies of the Dirac equation~(\ref{eq:Dirac}) (Exact) and Eq.~(\ref{eq:Sch1}) ($\varepsilon_{{1}/{M^2}}$) as well as their differences for four pseudospin partners.
  All units are in MeV.}
  \label{pt1}
  \begin{ruledtabular}
  \begin{tabular}{cccc}
  State & Exact & $\varepsilon_{{1}/{M^2}}$ & $\varepsilon_{{1}/{M^2}}-\mbox{Exact}$ \\
  \hline
  	$3p_{{3}/{2}}$ & \,\,\,$-7.6933$ & $-10.7831$ & $-3.2632$ \\
	$2f_{{5}/{2}}$ & \,\,\,$-8.7762$ & $-12.3894$ & $-3.7511$ \\
	$2f_{{7}/{2}}$ &      $-10.7584$ & $-13.6666$ & $-3.0079$ \\
	$1h_{{9}/{2}}$ &      $-14.1161$ & $-17.9759$ & $-3.8640$ \\
	$2d_{{5}/{2}}$ &      $-20.9981$ & $-23.9082$ & $-2.8937$ \\
	$1g_{{7}/{2}}$ &      $-24.7003$ & $-28.1825$ & $-3.3606$ \\
	$2p_{{3}/{2}}$ &      $-31.4007$ & $-34.0185$ & $-2.5216$ \\
	$1f_{{5}/{2}}$ &      $-34.7751$ & $-37.6507$ & $-2.6958$ \\
    \end{tabular}
  \end{ruledtabular}
\end{table}

By taking four pseudospin partners ($3p_{{3}/{2}}$, $2f_{{5}/{2}}$), ($2f_{{7}/{2}}$, $1h_{{9}/{2}}$), ($2d_{{5}/{2}}$, $1g_{{7}/{2}}$), and ($2p_{{3}/{2}}$, $1f_{{5}/{2}}$) as examples, the eigenenergies $\varepsilon_{{1}/{M^2}}$ obtained from solving Eq.~(\ref{eq:Sch1}) are shown in Table.~\ref{pt1}.
The results of the Dirac equation~(\ref{eq:Dirac}) labelled as Exact, which serve as benchmarks, are also shown in the table.
It is found that the eigenenergies $\varepsilon_{{1}/{M^2}}$ are all over bound by around $3$~MeV.

\begin{table}
  \caption{Contributions of operators $O_i$ in the $1/M^3$ order to the single-particle energies for the states $2d_{5/2}$, $1g_{7/2}$, $2f_{7/2}$, and $1h_{9/2}$. The last line shows the total contributions in this order.
  All units are in MeV.}
  \label{pt2}
  \begin{ruledtabular}
  \begin{tabular}{cccccc}
  $i$ & Operator & $2f_{{7}/{2}}$ & $1h_{{9}/{2}}$ & $2d_{{5}/{2}}$ & $1g_{{7}/{2}}$ \\
  \hline\vspace{0.3em}
  1 & $\displaystyle -\frac{1}{8M^3}p^4$ &                                    $-0.4783$ &  $-0.4144$ &	$-0.3370$ &	$-0.2707$ \\\vspace{0.3em}
  2 & $\displaystyle\frac{S^2}{2M^3}p^2$ &                          \,\,\,\,$3.1763$ &	\,\,\,\,$3.0608$ &\,\,\,\,$2.8875$ & \,\,\,\,$2.6501$\\\vspace{0.3em}
  3 & $\displaystyle\frac{-SS'}{M^3}\frac{\textrm{d}}{\textrm{d}r}$ &   \,\,\,\,$0.0517$ &	$-0.1752$ &	$-0.0291$ & $-0.1956$ \\\vspace{0.3em}
  4 & $\displaystyle -\frac{S\Sigma''}{4M^3} $    &              \,\,\,\,$0.0111$ &  \,\,\,\,$0.0301$ &	\,\,\,\,$0.0160$ &	\,\,\,\,$0.0285$ \\\vspace{0.3em}
  5 & $\displaystyle\frac{S\Delta'}{2M^3}\frac{\kappa}{r}$ &                  $-0.3138$ &	\,\,\,\,$0.5647$ &	$-0.2398$ &	\,\,\,\,$0.3942$\\\vspace{0.3em}
  6 & $\displaystyle -\frac{\Sigma'^2}{16M^3}$ &                              $-0.0007$ &	$-0.0006$ &	$-0.0006$ &	$-0.0005$\\\vspace{0.3em}
  7 & $\displaystyle\frac{\Sigma'\Delta'}{8M^3}$&                             $-0.0144$ &	$-0.0127$ &	$-0.0121$ &	$-0.0095$ \\\vspace{0.3em}
  Total &&                                                              \,\,\,\,$2.4319$ & 	\,\,\,\,$3.0527$ & \,\,\,\,$2.2849$ &	\,\,\,\,$2.5964$\\
  \end{tabular}
  \end{ruledtabular}
\end{table}

In the $1/M^3$ order, there are in total seven operators $O_i$ ($i = 1,\, 2,\, 3,\, \cdots,\, 7$) as shown in Eq.~(\ref{eq:p3}).
Their contributions to the single-particle energies are shown in Table.~\ref{pt2} for the states $2d_{5/2}$, $1g_{7/2}$, $2f_{7/2}$, and $1h_{9/2}$.
The contribution of each operator is calculated by the perturbation theory, i.e., $\varepsilon_i(k)=\langle k|O_i|k\rangle =\int \varphi_k^\ast O_i\varphi_k\,\textrm{d}\bm{r}$.

The operator $O_1$ corresponds to the relativistic correction to the kinetic energy, and its contributions are in general negative.
The operator $O_5$ is related to the spin-orbit interaction, so its contributions are also substantial, and they are negative (positive) for the spin-up states with $j=l+1/2$ (the spin-down states with $j=l-1/2$).
In contrast, the contributions from all the central terms $O_4$, $O_6$, and $O_7$ are relatively minor, and their influences can be ignored here.
Because of $S/M$ is approximately $1/3$, the biggest corrections in this order come from the operator $O_2$, which is related to the dynamical effects.
Its contributions are in general positive and as big as around $3$~MeV.
Finally, although the contributions of another dynamical term $O_3$ are not negligible, their specific values depend on the detailed structures of the operator as well as the single-particle wave functions.

\begin{table}
  \caption{Single-particle energies $\varepsilon_{{1}/{M^3}}$ obtained by the conventional SRG method with the perturbation corrections up to the $1/M^3$ order.
  The exact values and the differences between $\varepsilon_{{1}/{M^3}}$ and the exact ones are also shown.
  All units are in MeV.}
  \label{ptb3}
  \begin{ruledtabular}
  \begin{tabular}{cccc}
  State & Exact & $\varepsilon_{{1}/{M^3}}$ & $\varepsilon_{{1}/{M^3}} - \mbox{Exact}$ \\
  \hline
  	$3p_{{3}/{2}}$ & \,\,\,$-7.6933$ & \,\,\,$-8.1468$ & $-0.4536$ \\
	$2f_{{5}/{2}}$ & \,\,\,$-8.7762$ & \,\,\,$-9.3822$ & $-0.6059$ \\
	$2f_{{7}/{2}}$ &      $-10.7584$ & $-11.2347$ & $-0.4763$ \\
	$1h_{{9}/{2}}$ &      $-14.1161$ & $-14.9232$ & $-0.8071$ \\
	$2d_{{5}/{2}}$ &      $-20.9981$ & $-21.6233$ & $-0.6252$ \\
	$1g_{{7}/{2}}$ &      $-24.7003$ & $-25.5861$ & $-0.8858$ \\
	$2p_{{3}/{2}}$ &      $-31.4007$ & $-32.0691$ & $-0.6684$ \\
	$1f_{{5}/{2}}$ &      $-34.7751$ & $-35.6041$ & $-0.8290$ \\
    \end{tabular}
  \end{ruledtabular}
\end{table}

After adding the contributions in the $1/M^3$ order to the eigenenergies of Eq.~(\ref{eq:Sch1}), the single-particle energies $\varepsilon_{{1}/{M^3}}$ obtained by the conventional SRG method with the perturbation corrections up to the $1/M^3$ order are shown in Table.~\ref{ptb3}.
All four pseudospin partners are still over bound.
Although they are reduced to the values between $0.5$~MeV and $1$~MeV for most states, the differences between $\varepsilon_{{1}/{M^3}}$ and the exact ones still cannot be ignored.
Therefore, it is necessary to further calculate the corrections from the ${1}/{M^4}$ order.

\begin{table*}
  \caption{Same as Table~\ref{pt2}, but for the operators $O_i$ in the $1/M^4$ order.}\label{pt3}
  \begin{ruledtabular}
  \begin{tabular}{ccccccc}
   $i$ & Operator & $2f_{{7}/{2}}$ & $1h_{{9}/{2}}$ & $2d_{{5}/{2}}$ & $1g_{{7}/{2}}$ \\
  \hline\vspace{0.3em}
  8 & $\displaystyle\frac{3S}{8M^4}p^4$&$-0.5338$ &	$-0.4579$	&$-0.3878$ &	$-0.3099$ 	\\\vspace{0.3em}
  9 & $\displaystyle -\frac{3S'}{4M^4}p^2\frac{\textrm{d}}{\textrm{d}r}$&\,\,\,\,$0.0464$ &	\,\,\,\,$0.0514$  	&\,\,\,\,$0.0474$ 	&\,\,\,\,$0.0413$\\\vspace{0.3em}
  10 & $\displaystyle\frac{3\Delta'}{16M^4}\frac{\kappa}{r}p^2$ & \,\,\,\,$0.0233$ 	&$-0.0363$ &	\,\,\,\,$0.0121$ &	$-0.0177$\\\vspace{0.3em}
  11 & $\displaystyle -\frac{3\Sigma''}{16M^4}p^2$&$-0.0020$ &	$-0.0043$  &	$-0.0021$ &	$-0.0030$ \\\vspace{0.3em}
  12 &$\displaystyle -\frac{9S''}{16M^4}p^2$& $-0.0320$ &	$-0.0696$  &	$-0.0349$ &	$-0.0496$ \\\vspace{0.3em}
  13 &$\displaystyle -\frac{S^3}{2M^4}p^2$& \,\,\,\,$1.1399$ &	\,\,\,\,$1.0671$  &	\,\,\,\,$1.0466$ &	\,\,\,\,$0.9472$\\\vspace{0.3em}
  14 &$\displaystyle\frac{3\Delta'}{16M^4r}\frac{\kappa}{r}\frac{\textrm{d}}{\textrm{d}r}$& \,\,\,\,$0.0000$ &	\,\,\,\,$0.0013$  &	$-0.0003$ &	\,\,\,\,$0.0011$ \\\vspace{0.3em}
  15 & $\displaystyle -\frac{3\Delta''}{16M^4}\frac{\kappa}{r}\frac{\textrm{d}}{\textrm{d}r}$&$-0.0067$ &	\,\,\,\,$0.0005$ &	$-0.0036$ &	$-0.0019$ \\\vspace{0.3em}
  16 &$\displaystyle\frac{3S'''}{16M^4}\frac{\textrm{d}}{\textrm{d}r}$& \,\,\,\,$0.0032$ &	\,\,\,\,$0.0084$  &	\,\,\,\,$0.0073$ &	\,\,\,\,$0.0062$ \\\vspace{0.3em}
  17 &$\displaystyle\frac{3\Sigma'''}{16M^4}\frac{\textrm{d}}{\textrm{d}r}$& \,\,\,\,$0.0006$ &	\,\,\,\,$0.0016$  &	\,\,\,\,$0.0013$ &	\,\,\,\,$0.0011$\\\vspace{0.3em}
  18 & $\displaystyle\frac{3S^2S'}{2M^4}\frac{\textrm{d}}{\textrm{d}r}$& \,\,\,\,$0.0417$ &	$-0.0528$  &	\,\,\,\,$0.0117$ &	$-0.0693$ \\\vspace{0.3em}
  19 & $\displaystyle -\frac{3\Sigma'}{32M^4r}\frac{\kappa(\kappa+1)}{r^2}$& $-0.0001$ &	$-0.0003$  &	\,\,\,\,$0.0000$ &	$-0.0002$ \\\vspace{0.3em}
  20& $\displaystyle\frac{9S'}{8M^4r}\frac{\kappa(\kappa+1)}{r^2}$&       \,\,\,\,$0.0056$ &	\,\,\,\,$0.0179$  &	\,\,\,\,$0.0027$ &	\,\,\,\,$0.0105$ \\\vspace{0.3em}
  21 &$\displaystyle -\frac{3\Delta'}{16M^4r^2}\frac{\kappa}{r}$& $-0.0006$ &	\,\,\,\,$0.0009$ 	 &	$-0.0004$ &	\,\,\,\,$0.0006$ \\\vspace{0.3em}
  22& $\displaystyle\frac{3\Sigma''}{32M^4}\frac{\kappa(\kappa+1)}{r^2}$ & \,\,\,\,$0.0002$ &	\,\,\,\,$0.0014$  &	\,\,\,\,$0.0001$ &	\,\,\,\,$0.0009$ \\\vspace{0.3em}
  23 & $\displaystyle\frac{3S''}{8M^4}\frac{\kappa(\kappa+1)}{r^2}$&   \,\,\,\,$0.0038$ &	\,\,\,\,$0.0295$  &    \,\,\,\,$0.0030$ &	\,\,\,\,$0.0193$ \\\vspace{0.3em}
  24 &$\displaystyle\frac{3\Delta''}{16M^4r}\frac{\kappa}{r}$& \,\,\,\,$0.0011$ &	$-0.0045$  &	\,\,\,\,$0.0014$ &	$-0.0035$ \\\vspace{0.3em}
  25 & $\displaystyle -\frac{3\Delta'''}{32M^4}\frac{\kappa}{r}$& \,\,\,\,$0.0062$ &	\,\,\,\,$0.0017$ & 	\,\,\,\,$0.0029$ &	\,\,\,\,$0.0036$ \\\vspace{0.3em}
  26 & $\displaystyle -\frac{3S^2\Delta'}{4M^4}\frac{\kappa}{r}$& $-0.1170$ &	\,\,\,\,$0.2510$ 	&	$-0.0969$ &	\,\,\,\,$0.1816$ \\\vspace{0.3em}
  27 & $\displaystyle\frac{9\Sigma''''}{128M^4}$&  $-0.0004$ &	$-0.0012$ 	 &	$-0.0010$ &	$-0.0009$ \\\vspace{0.3em}
  28 & $\displaystyle\frac{3S^2\Sigma''}{8M^4}$&  \,\,\,\,$0.0058$& 	\,\,\,\,$0.0150$	&\,\,\,\,$0.0079$ 	&\,\,\,\,$0.0143$ \\\vspace{0.3em}
  29 & $\displaystyle\frac{3S\Sigma'^2}{16M^4}$&  $-0.0004$ &	$-0.0005$  &	$-0.0004$ 	&$-0.0004$ \\\vspace{0.3em}
  30 & $\displaystyle -\frac{3S\Sigma'\Delta'}{8M^4}$&$-0.0088$ &	$-0.0092$ & 	$-0.0081$ &	$-0.0072$ \\\vspace{0.3em}
  Total & &\,\,\,\,$0.5760$ &	\,\,\,\,$0.8113$	&\,\,\,\,$0.6089$ &	\,\,\,\,\,$0.7641$ \\
  \end{tabular}
  \end{ruledtabular}
\end{table*}

In the $1/M^4$ order, there are in total twenty three operators $O_i$ ($i = 8,\, 9,\, 10,\, \cdots,\, 30$) as shown in Eq.~(\ref{eq:order4p}), and their contributions to the single-particle energies calculated by the perturbation theory are shown in Table.~\ref{pt3}.

The operator $O_8$ is also related to the relativistic correction to the kinetic energy, and its contributions are negative and comparable to those of $O_1$ in the previous order.
Due to the same reason, the biggest corrections in this order come from the dynamical term $O_{13}$, and its contributions are positive and around $1/3$ of those of $O_2$.
Another dynamical term $O_9$ provides small positive contributions as well.
The operators $O_{26}$ and $O_{10}$ are related to the spin-orbit interaction, and thus the signs of their contributions depend on the states.
The operators $O_{14}$, $O_{15}$, $O_{21}$, $O_{24}$, and $O_{25}$ are also related to the spin-orbit interaction, but their contributions can be ignored.
The operators $O_{19}$, $O_{20}$, $O_{22}$, and $O_{23}$ are related to the centrifugal barrier, and the operators $O_{27}$, $O_{28}$, $O_{29}$, and $O_{30}$ are related to the central force or mean-field terms, but their contributions are all relatively minor.
Finally, the origins of the other small terms are more complicated, and their specific contributions depend on the detailed structures.

\begin{table}
  \caption{Same as Table.~\ref{ptb3}, but for the single-particle energies $\varepsilon_{{1}/{M^4}}$ with the perturbation corrections up to the $1/M^4$ order.}
  \label{pt4}
  \begin{ruledtabular}
  \begin{tabular}{cccc}
  State & Exact & $\varepsilon_{{1}/{M^4}}$ & $ \varepsilon_{{1}/{M^4}} - \mbox{Exact}$\\
  \hline
    $3p_{{3}/{2}}$ & \,\,\,$-7.6933$  & \,\,\,$-7.5200$  & \,\,\,$0.1733$\\
    $2f_{{5}/{2}}$ & \,\,\,$-8.7762$  & \,\,\,$-8.6384$  & \,\,\,$0.1379$\\
    $2f_{{7}/{2}}$ & $-10.7584$ & $-10.6587$ & \,\,\,$0.0997$\\
    $1h_{{9}/{2}}$ & $-14.1161$ & $-14.1119$ & \,\,\,$0.0042$\\
    $2d_{{5}/{2}}$ & $-20.9981$ & $-21.0144$ & $-0.0163$\\
    $1g_{{7}/{2}}$ & $-24.7003$ & $-24.8220$ & $-0.1216$\\
    $2p_{{3}/{2}}$ & $-31.4007$ & $-31.4969$ & $-0.0962$\\
    $1f_{{5}/{2}}$ & $-34.7751$ & $-34.9549$ & $-0.1798$\\
  \end{tabular}
  \end{ruledtabular}
\end{table}

After adding both the contributions in the $1/M^3$ and $1/M^4$ orders to the eigenenergies of Eq.~(\ref{eq:Sch1}), the single-particle energies $\varepsilon_{{1}/{M^4}}$ obtained by the conventional SRG method with the perturbation corrections up to the $1/M^4$ order are shown in Table.~\ref{pt4}.
Now the deeply bound states are still slightly over bound, while the weakly bound states become under bound.
It is clear that the differences between the single-particle energies $\varepsilon_{{1}/{M^4}}$ and the exact ones are less than $0.2$~MeV for all states, and even less than $0.1$~MeV for some states.

\begin{figure}
  \includegraphics[width=0.45\textwidth]{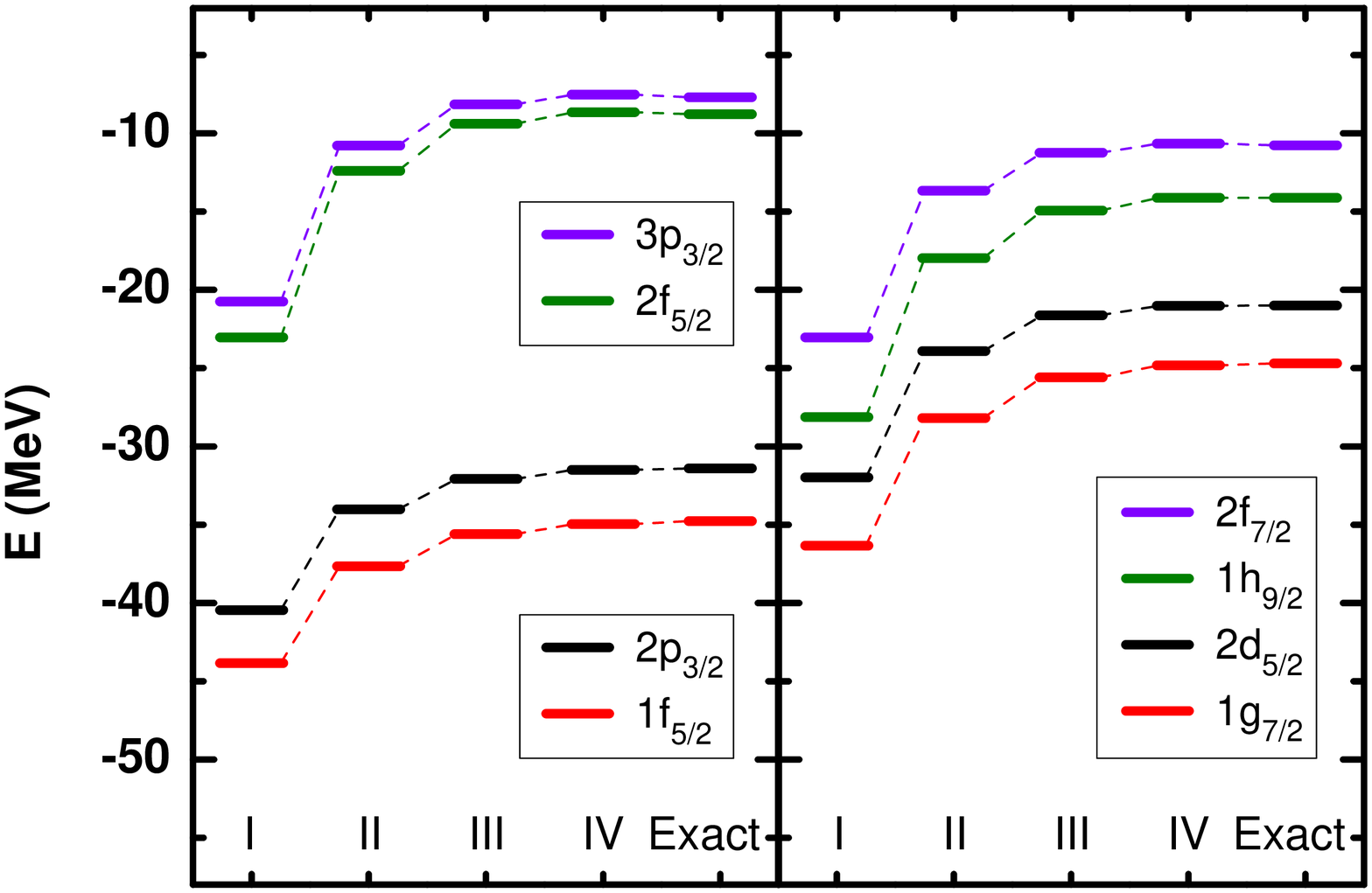}
  \caption{(Color online) Energy spectrum of $\mathcal{H}$ for the four pseudospin partners calculated by the conventional SRG method.
  The first, second, third, and fourth columns show the single-particle energies up to the $1/M$, $1/M^2$, $1/M^3$, and $1/M^4$ orders, respectively.
  The last column labelled as Exact shows the eigenenergies of the Dirac equation~(\ref{eq:Dirac}).}\label{fig:1}
\end{figure}

In order to show the results concisely, the energy spectrum of $\mathcal{H}$ for the four pseudospin partners calculated by the conventional SRG method is shown in Fig.~\ref{fig:1}.
The first, second, third, and fourth columns show the single-particle energies up to the $1/M$, $1/M^2$, $1/M^3$, and $1/M^4$ orders, respectively.
The last column labelled as Exact shows the eigenenergies of the Dirac equation~(\ref{eq:Dirac}).

For the discussions on the pseudospin symmetry, it is found that among the perturbation corrections in the $1/M^3$ order, the spin-orbit term $O_5$ and the relativistic correction to the kinetic energy $O_1$ reduce the pseudospin-orbit splittings, while the dynamical terms $O_2$ and $O_3$ increase the splittings.
It is consistent with the findings in Ref.~\cite{Guo2012Phys.Rev.C85.021302}.
Furthermore, among the perturbation corrections in the $1/M^4$ order, on the one hand, the dominant spin-orbit and relativistic correction terms are $O_{26}$ and $O_{8}$, respectively, and they reduce further the pseudospin-orbit splittings substantially;
on the other hand, the dominant dynamical terms $O_{13}$ and $O_{18}$ increase the splittings.
This supports the main conclusion in Ref.~\cite{Guo2012Phys.Rev.C85.021302}.

\subsection{Single-particle energy with reconstituted SRG}\label{sec:IIIB}

\begin{table}
  \caption{Eigenenergies of the Dirac equation~(\ref{eq:Dirac}) (Exact) and Eq.~(\ref{eq:Sch2}) ($\varepsilon_{{1}/{{M^\ast}^2}}$) as well as their differences for four pseudospin partners.
  All units are in MeV.}
  \label{nt1}
  \begin{ruledtabular}
  \begin{tabular}{cccc}
  State & Exact & $\varepsilon_{{1}/{{M^\ast}^2}}$ & $\varepsilon_{{1}/{{M^\ast}^2}} - \mbox{Exact}$\\
  \hline
  	$3p_{{3}/{2}}$ &\,\,\,$-7.6933$ & \,\,\,$-6.2635$ & $1.4297$\\
	$2f_{{5}/{2}}$ & \,\,\,$-8.7762$ & \,\,\,$-7.2580$ & $1.5183$\\
	$2f_{{7}/{2}}$ & $-10.7584$ & \,\,\,$-9.3517$ & $1.4067$\\
	$1h_{{9}/{2}}$ & $-14.1161$ & $-12.7160$ & $1.4000$\\
	$2d_{{5}/{2}}$ & $-20.9981$ & $-19.9176$ & $1.0805$\\
	$1g_{{7}/{2}}$ & $-24.7003$ & $-23.7312$ & $0.9691$\\
    $2p_{{3}/{2}}$ & $-31.4007$ & $-30.6919$ & $0.7088$\\
	$1f_{{5}/{2}}$ & $-34.7751$ & $-34.1796$ & $0.5955$\\
  \end{tabular}
  \end{ruledtabular}
\end{table}

Let us then investigate the speed of convergence of the newly proposed reconstituted SRG method.
First of all, by taking the same four pseudospin partners as examples, the eigenenergies obtained from solving Eq.~(\ref{eq:Sch2}) are shown in Table.~\ref{nt1}.
The results of the Dirac equation~(\ref{eq:Dirac}) labelled as Exact also serve as benchmarks in the table.
Different from the conventional SRG method, for four pseudospin partners, they are all under bound, and the differences between the eigenenergies $\varepsilon_{{1}/{{M^\ast}^2}}$ and the exact ones are around $1$~MeV, which are already reduced a lot compared with the case of $\varepsilon_{{1}/{M^2}}$.

\begin{table}
  \caption{Contributions of operators $\tilde O_i$ in the $1/{M^\ast}^3$ order to the single-particle energies for the states $2d_{5/2}$, $1g_{7/2}$, $2f_{7/2}$, and $1h_{9/2}$. The last line shows the total contributions in this order.
  All units are in MeV.}\label{nt2}
  \begin{ruledtabular}
  \begin{tabular}{ccccccc}
  $i$ & Operator & $2f_{{7}/{2}}$ & $1h_{{9}/{2}}$ & $2d_{{5}/{2}}$ & $1g_{{7}/{2}}$ \\
  \hline\vspace{0.3em}
  1 &$\displaystyle -p^2\frac{1}{8{M^\ast}^3}p^2$& $-1.4605$ &	$-1.2577$  &	$-1.1138$ &	$-0.8952$ \\\vspace{0.3em}
  2 & $\displaystyle\frac{\Sigma'^2}{16{M^\ast}^3}$ & \,\,\,\,$0.0015$ &	\,\,\,\,$0.0018$  &	\,\,\,\,$0.0015$ &	\,\,\,\,$0.0014$\\\vspace{0.3em}
  3 & $\displaystyle -\frac{S'\Sigma'}{4{M^\ast}^3}$&$-0.0320$ &	$-0.0382$  &	$-0.0317$ &   $-0.0302$ \\\vspace{0.3em}
  Total && $-1.4911$ &	$-1.2941$  &	$-1.1441$ 	&$-0.9240$ \\
  \end{tabular}
  \end{ruledtabular}
\end{table}

In the $1/{M^\ast}^3$ order, there are only three operators $\tilde{O}_i$ ($i = 1,\, 2,\, 3$) as shown in Eq.~(\ref{eqn3}).
Their contributions to the single-particle energies for the states $2d_{5/2}$, $1g_{7/2}$, $2f_{7/2}$, and $1h_{9/2}$ are shown in Table.~\ref{nt2}.
The contribution of each operator is also calculated by the perturbation theory, i.e., $\tilde{\varepsilon}_i(k)=\langle \tilde{k}|\tilde{O}_i|\tilde{k}\rangle =\int \tilde{\varphi}_k^\ast \tilde{O}_i\tilde{\varphi}_k\,\textrm{d}\bm{r}$.
The dominant corrections come from the relativistic correction to the kinetic energy $\tilde{O}_1$.
They are in general negative, and their values are approximately as three times as those of ${O}_1$ in the conventional SRG, mainly because of the factor $(M/M^\ast)^3$.
On the other hand, the essential corrections $O_2$, $O_3$, and $O_5$ in the conventional SRG have been absorbed in the lower orders in the present reconstituted SRG method.

\begin{table}
  \caption{Single-particle energies $\varepsilon_{{1}/{{M^\ast}^3}}$ obtained by the reconstituted SRG method with the perturbation corrections up to the $1/{{M^\ast}^3}$ order.
  The exact values and the differences between $\varepsilon_{{1}/{{M^\ast}^3}}$ and the exact ones are also shown.
  All units are in MeV.}
  \label{nt3}
  \begin{ruledtabular}
  \begin{tabular}{cccc}
  State & Exact & $\varepsilon_{{1}/{{M^\ast}^3}}$ & $\epsilon_{{1}/{{M^\ast}^3}} - \mbox{Exact}$\\
  \hline
  $3p_{{3}/{2}}$ & \,\,\,$-7.6933$ & \,\,\,$-7.7167$ & $-0.0235$\\
  $2f_{{5}/{2}}$ & \,\,\,$-8.7762$ & \,\,\,$-8.7270$ & \,\,\,\,$0.0492$\\
  $2f_{{7}/{2}}$ & $-10.7584$ & $-10.8428$ & $-0.0844$\\
  $1h_{{9}/{2}}$ & $-14.1161$ & $-14.0101$ & \,\,\,\,$0.1060$\\
  $2d_{{5}/{2}}$ & $-20.9981$ & $-21.0617$ & $-0.0635$\\
  $1g_{{7}/{2}}$ & $-24.7003$ & $-24.6552$ & \,\,\,\,$0.0451$\\
  $2p_{{3}/{2}}$ & $-31.4007$ & $-31.4342$ & $-0.0335$\\
  $1f_{{5}/{2}}$ & $-34.7751$ & $-34.7598$ & \,\,\,\,$0.0153$\\
  \end{tabular}
  \end{ruledtabular}
\end{table}

After adding the contributions in the $1/{M^\ast}^3$ order to the eigenenergies of Eq.~(\ref{eq:Sch2}), the single-particle energies $\varepsilon_{{1}/{{M^\ast}^3}}$ obtained by the reconstituted SRG method with the perturbation corrections up to the $1/{M^\ast}^3$ order are shown in Table.~\ref{nt3}.
Now the spin-up states are slightly over bound, while the spin-down states are slightly under bound.
It is clear that the differences between $\varepsilon_{{1}/{{M^\ast}^3}}$ and the exact ones are in general less than $0.1$~MeV, and for some states the differences are even negligible.
Therefore, it is not necessary to go to higher orders in this novel expansion.

\begin{figure}
  \includegraphics[width=0.45\textwidth]{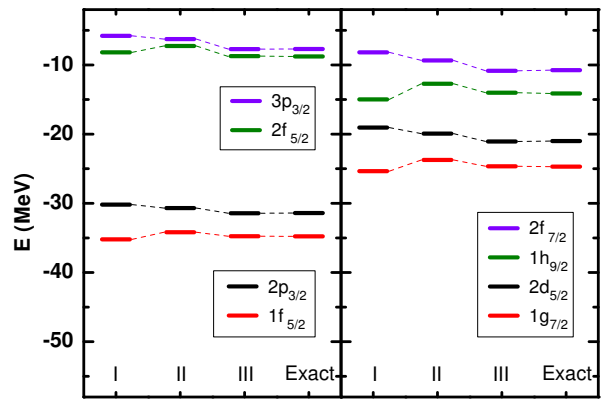}
  \caption{(Color online) Energy spectrum of $\tilde{\mathcal{H}}$ for the four pseudospin partners calculated by the reconstituted SRG method.
  The first, second, and third columns show the single-particle energies up to $\tilde{\mathcal{H}}_1$, $\tilde{\mathcal{H}}_2$, and $\tilde{\mathcal{H}}_3$, respectively.
  The last column labelled as Exact shows the eigenenergies of the Dirac equation~(\ref{eq:Dirac}).}\label{fig:2}
\end{figure}

In order to show the results concisely, the energy spectrum of $\tilde{\mathcal{H}}$ for the four pseudospin partners calculated by the reconstituted SRG method is shown in Fig.~\ref{fig:2}.
The first, second, and third columns show the single-particle energies up to $\tilde{\mathcal{H}}_1$, $\tilde{\mathcal{H}}_2$, and $\tilde{\mathcal{H}}_3$, respectively.
The last column labelled as Exact shows the eigenenergies of the Dirac equation~(\ref{eq:Dirac}).

\subsection{Single-particle density}\label{sec:IIIC}

The single-particle density is another important physical quantity, in particular, in the context of DFT.
Moreover, as the Hamiltonian has been transformed by unitary operators in SRG, the single-particle wave functions are transformed correspondingly, but the single-particle densities remain the same.
Therefore, it is also worthwhile to investigate the single-particle densities $\rho(\bm{r})$.

First, the single-particle densities $\rho(\bm{r})=\varphi^\ast(\bm{r})\varphi(\bm{r})$ ($\rho(\bm{r})=\tilde\varphi^\ast(\bm{r})\tilde\varphi(\bm{r})$) are calculated from the conventional (reconstituted) SRG method by using the solutions of Eq.~(\ref{eq:Sch1}) (Eq.~(\ref{eq:Sch2})).
Then, according to the perturbation theory, the wave functions are corrected as
\begin{subequations}
\begin{align}
\varphi_k^{(1)}&=\varphi_k+\sum_{n\neq k}a_{nk}\varphi_n,\\
\tilde{\varphi}_k^{(1)}&=\tilde{\varphi}_k+\sum_{n\neq k}\tilde{a}_{nk}\tilde{\varphi}_n,
\end{align}
\end{subequations}
and here
\begin{subequations}
\begin{align}
a_{nk}&=\frac{\langle n | \mathcal{H}_3 | k \rangle}{\varepsilon_k-\varepsilon_n},\\
\tilde{a}_{nk}&=\frac{\langle n | \tilde{\mathcal{H}}_3 | k \rangle}{\tilde{\varepsilon}_k-\tilde{\varepsilon}_n},
\end{align}
\end{subequations}
to the $1/M^3$ and $1/{M^\ast}^3$ orders, respectively.
Finally, the wave functions have been normalized after the perturbation.

\begin{figure*}
  \includegraphics[width=0.8\textwidth]{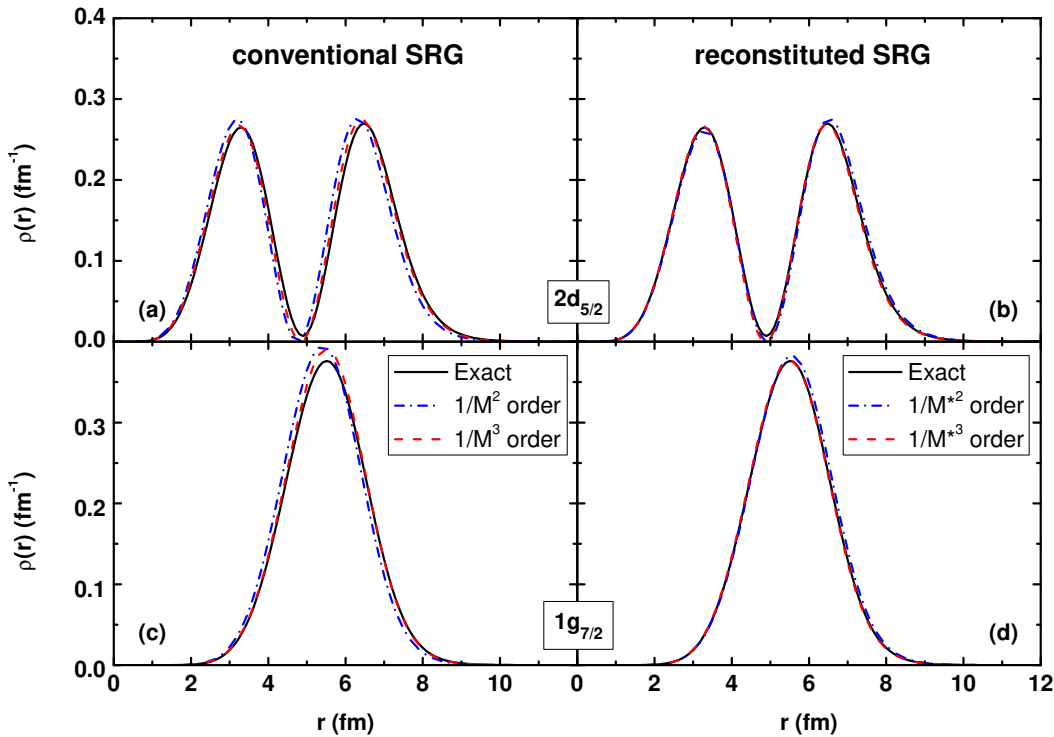}
  \caption{(Color online) Single-particle densities for the $2d_{5/2}$ (upper panels) and $1g_{7/2}$ (lower panels) states.
  The solutions of the Dirac equation~(\ref{eq:Dirac}) are labelled as Exact with the solid lines.
  The results by the conventional (reconstituted) SRG method are shown in the left (right) panels, where the solutions of Eq.~(\ref{eq:Sch1}) (Eq.~(\ref{eq:Sch2})) and those with the perturbation corrections up to the $1/M^3$ ($1/{M^\ast}^3$) order are shown with the dash-dotted and dashed lines, respectively.}
  \label{fig:3}
\end{figure*}

In Fig.~\ref{fig:3}, the single-particle densities for the $2d_{5/2}$ and $1g_{7/2}$ states are shown, where the results by the conventional and reconstituted SRG methods are compared with the benchmark solutions of the Dirac equation~(\ref{eq:Dirac}).
For the conventional method, compared with the exact densities, even with the corrections up to the $1/M^3$ order, obvious differences can still be seen.
In contrast, for the reconstituted method, the direct results in the $1/{M^\ast}^2$ order have improved the conventional ones.
After considering the corrections in the next order, the single-particle densities are almost identical to the exact ones.

\section{Summary and Perspectives}\label{sec:IV}

By following the conventional SRG method, we have successfully worked out the solutions of the $1/M^4$ order for the first time.
On the one hand, the investigation for the higher order has successfully verified the convergence of the conventional method.
On the other hand, the numerical results of the contributions in the $1/M^3$ and $1/M^4$ orders also support that the pseudospin splittings are added by the dynamical terms and are reduced by the spin-orbit interactions as well as the relativistic corrections to the kinetic energy.

With the re-summation of the operators, the reconstituted SRG method has been proposed. By replacing the bare mass of nucleon by the Dirac mass, the terms that belong to the same families as those in the lower orders are also summed to the infinite order.
The superiority of the reconstituted SRG method has been shown in two aspects. One is the single-particle energy. The speed of convergence with the SRG expansion has been improved significantly.
For instance, the results of the reconstituted SRG expansion up to the $1/{M^\ast}^3$ order are better than those obtained from the conventional SRG expansion up to the $1/M^4$ order.
The other one is the single-particle density. After considering the corrections from the $1/{M^\ast}^3$ order in the reconstituted SRG, the single-particle densities are almost identical to the exact ones.

For the future studies, this novel non-relativistic expansion method establishes a potential bridge between the relativistic and non-relativistic density functional theories.

\begin{acknowledgments}

The authors are grateful to Dr.~Tetsuo Hatsuda for the helpful discussions on the Hermitian properties of the Hamiltonian in the reconstituted SRG method.
Y.G. also acknowledges the financial supports from University of Science and Technology of China and the hospitality of RIKEN.
This work was partially supported by the JSPS Grant-in-Aid for Early-Career Scientists under Grant No.~18K13549 and the JSPS-NSFC Bilateral Program for Joint Research Project on Nuclear mass and life for unravelling mysteries of the $r$-process.

\end{acknowledgments}


%

\end{CJK*}
\end{document}